\documentclass[fleqn,12pt,twoside]{article}
\usepackage{espcrc1}

\usepackage{graphicx}
\usepackage[figuresright]{rotating}

\newcommand{\AmS}{{\protect\the\textfont2
  A\kern-.1667em\lower.5ex\hbox{M}\kern-.125emS}}

\def\al{\alpha}

\def\be{\begin{equation}}
\def\ee{\end{equation}}
\def \ba  {\begin{eqnarray}}
\def \ea  {\end{eqnarray}}
\def\bea{\begin{eqnarray}}
\def\eea{\end{eqnarray}}

\def \as {\relax\ifmmode\alpha_s\else{$\alpha_s${ }}\fi}

\def \al #1 {\frac {\as({#1})}{\pi} }
\def \ds #1 {\ooalign{$\hfil/\hfil$\crcr$#1$}}

\def \GeV {\mbox{GeV}}

\def\qt{Q_T}

\title{Phenomenological studies in QCD resummation}

\author{A. Kulesza\address[BNLHE]{Department of Physics, Brookhaven 
National Laboratory, Upton, NY 11973, U.S.A.}\thanks{A.K.\ was supported 
by the U.S. Department of Energy (contract number DE-AC02-98CH10886).},
G. Sterman\address[YITP]{C.N.\ Yang Institute for Theoretical Physics, 
          SUNY Stony Brook, Stony Brook, New York 11794 -- 3840, U.S.A.}\thanks{The work of G.S.\ was supported in part by the National
Science Foundation, grants PHY9722101 and PHY0098527.} and
        W. Vogelsang\address[RBRC]{RIKEN-BNL Research Center and 
         Nuclear Theory, Brookhaven National Laboratory, Upton, NY 11973, 
       U.S.A.}\thanks{Invited talk presented at the ``XVI International 
Conference on Particles and Nuclei (PaNic02)'', Osaka, Japan, September 2002.
W.V.\ is grateful to RIKEN, Brookhaven National Laboratory and the U.S.
Dept. of Energy (contract number DE-AC02-98CH10886) for 
providing the facilities essential for the completion of this work.}}

\begin{document}

\maketitle

\vspace*{-9.6cm}
\begin{flushright}
BNL-HET-03/2 \\
BNL-NT-02/28 \\
RBRC-302 \\
YITP-SB-03/5 \\
December 2002 
\end{flushright}

\vspace*{5.9cm}

\begin{abstract}
We study applications of QCD soft-gluon resummations to electroweak
annihilation cross sections. We focus on a formalism that
allows to resum logarithmic corrections arising
near partonic threshold and at small transverse momentum simultaneously. 
\end{abstract}

\vspace*{-1mm}
\section{INTRODUCTION}

\vspace*{-2mm}
When probed near an exclusive boundary of phase space,
perturbative partonic hard-scatt\-ering cross sections
for electroweak-boson ($\gamma^*$, W, Z, H) production 
acquire large logarithmic corrections arising from
incomplete cancellations of soft-gluon effects 
between virtual and real diagrams.
The two prominent examples are {\em threshold} and {\em recoil} 
corrections. The former are of the form $\as^n \ln^{2n-1}(1-z)/(1-z)$ 
and become large when the partonic c.m. energy approaches the invariant
mass $Q$ of the produced boson, $z=Q^2/\hat s \rightarrow 1$. 
The recoil corrections, in turn, are of the form $\as^n
\ln^{2n-1}(Q^2/Q_T^2)$ and grow large if the transverse momentum
carried by the produced boson is very small, $Q_T \ll Q$. Therefore, 
sufficiently close to the phase-space boundary, i.e. in the limit of soft
and/or collinear radiation, fixed-order perturbation theory is bound to fail.
A proper treatment of the cross section requires resummation of the 
logarithmic corrections to all orders. The techniques for this 
are well established in both the threshold~\cite{Sthr,CTthr} 
and in the recoil~\cite{DDT,PP,AEGM,CSS} cases.

Resummation of recoil and threshold corrections, however, is known
to lead to opposite effects -- suppression and enhancement of the partonic 
cross section, respectively. A full analysis of soft gluon effects in
transverse momentum distributions $d\sigma/dQ^2\,dQ_T^2$ should therefore,
if possible, take both types of corrections simultaneously into account. 
A joint treatment of the threshold and recoil 
corrections was proposed in~\cite{LSVprl,LSV}.  It relies on a
novel refactorization of short-distance and long-distance physics at fixed
transverse momentum and energy~\cite{LSV}. Similarly to standard threshold and 
recoil resummations, exponentiation of logarithmic corrections occurs in the
impact parameter $b$ space, Fourier-conjugated to transverse momentum $Q_T$
space, and Mellin-$N$ moment space, conjugated to $z$ space. This time 
both transforms are present, resulting in a final expression which
obeys energy and transverse-momentum conservation. Consequently, 
phenomenological evaluation of the joint resummation expressions requires 
prescriptions for inverse transforms from both $N$ and $b$ 
spaces. This issue is also closely tied to specifying the border between 
resummed perturbation theory and the nonperturbative regime, through
analysis of the nonperturbative effects implied by the resummation
formula itself. Moreover, to fully define the expressions a 
procedure for matching between the fixed-order and the resummed result 
needs to be specified. A full phenomenological study of the joint resummation 
formalism as applied to vector boson production was undertaken in~\cite{KSV}.
The formalism may also be applied to Higgs production via gluon-gluon 
fusion~\cite{KSVHiggs}. In this case the Higgs-gluon interaction proceeds
through a top quark loop and may, for $m_t>m_h$, be replaced~\cite{eff} 
by a simple effective $ggh$ vertex. In the following we will briefly 
discuss our results for joint resummation as applied to eletroweak-boson 
production.

\section{THE JOINTLY RESUMMED CROSS SECTION}

In the  framework of joint resummation, the resummed electroweak 
annihilation cross section has the following form~\cite{LSV,KSV}:
\bea
\label{crsec}
     \frac{d\sigma_{AB}^{\rm res}}{dQ^2\,dQ_T^2}
     &=&    \sum_a \sigma_{a}^{(0)}(Q^2)\,
\int_{C_N}\, \frac{dN}{2\pi i} \,\tau^{-N}\;    \int \frac{d^2b}{(2\pi )^2} \,
e^{i{\vec{Q}_T}\cdot {\vec{b}}}\, \nonumber \\
&\times&    {\cal C}_{a/A}(Q,b,N,\mu,\mu_F )\;
      \exp\left[ \,E_{a\bar a}^{\rm PT} (N,b,Q,\mu)\,\right] \;
      {\cal C}_{\bar{a}/B}(Q,b,N,\mu,\mu_F) \; ,
\eea
where $\sigma_a^{(0)}(Q^2)$ denotes a perturbative normalization that only
depends on the large invariant mass $Q$ of the produced 
boson~\cite{KSV,KSVHiggs}. We have 
defined $\tau=Q^2/S$. The flavor-diagonal exponent 
$E_{a\bar a}^{\rm PT}$ was derived in~\cite{KSV} to next-to-leading 
logarithmic (NLL) accuracy:
\bea
\label{jointsud}
E_{a\bar a}^{\rm PT} (N,b,Q,\mu,\mu_F) &=&
-\int_{Q^2/\chi^2}^{Q^2} {d k_T^2 \over k_T^2} \;
\left[ A_a(\as(k_T))\,
\ln\left( {Q^2 \over k_T^2} \right) + B_a(\as(k_T))\right] \,.
\eea
It has the classic form of the Sudakov exponent in the 
recoil-resummed $\qt$ distribution for
electroweak annihilation, with the $A$ and $B$ functions defined
as perturbative series in $\as$~\cite{DDT,PP,AEGM,CSS}. 
The quantity $\chi(N,b)$ organizes the logarithms of $N$ and $b$ in 
joint resummation~\cite{KSV}:
\be
\label{chinew}
\chi(\bar{N},\bar{b})=\bar{b} + \frac{\bar{N}}{1+\,\bar{b}/
4\bar{N}}\; ,
\ee
where we define
$\bar{N} = N{\rm e^{\gamma_E}}$,
$\bar{b}= b Q {\rm e^{\gamma_E}}/2$, with $\gamma_E$ the Euler 
constant. With this choice for $\chi(\bar{N},\bar{b})$
the LL and NLL terms are correctly reproduced in the threshold limit, 
$N \rightarrow \infty$ (at fixed $b$), and in the recoil limit 
$b \rightarrow \infty$ (at fixed $N$). 

The coefficients 
in the expansions of the functions in~(\ref{jointsud})
are the same as in the pure $\qt$
resummation and are known from comparison with fixed-order 
calculations~\cite{Kodaira:1982nh,Catani:1988vd,Davies:1984hs,dFG} for
both vector-boson and Higgs production.  At NLL only
$A^{(1)},\ B^{(1)}$ and $A^{(2)}$ contribute,  
\bea
A_a^{(1)} &=&  C_a\;, \;\;\;\;\; A_a^{(2)} \;=\; \frac{C_a}{2} \left[
C_A \left( \frac{67}{18}-\frac{\pi^2}{6} \right) -\frac{10}{9}T_R
N_F\right]\;, \;\;\;\;\;\; (C_q=C_F\;,\;\;C_g=C_A) \; , \nonumber \\
B_q^{(1)} &=&-\frac{3}{2} C_F\;,\;\;\;\;\;\;
B_g^{(1)}\;=\;-\frac{1}{6} \left( 11 C_A - 4 T_R N_F \right)\; .
\eea
The second-order term $B_g^{(2)}$ contributes only at 
NNLL level. It was noted in a previous study on $Q_T$ resummation for
Higgs production~\cite{HN} that the contribution from $B_g^{(2)}$ is actually 
numerically rather significant due to the size of $C_A$, 
and we therefore include it in our study
despite the fact that it is subleading to our analysis. We also note that 
there is an interplay~\cite{CdFG} between $B_g^{(2)}$, the  function 
$\sigma_a^{(0)}(Q^2)$
above, and the coefficients ${\cal C}_{a/H}$ to be specified below; 
for details, see~\cite{CdFG}. We will use in the case of Higgs 
production~\cite{dFG,CdFG}
\be
B_g^{(2)}= 
C_A^2\left(-{4 \over 3}+{11 \over 36}\pi^2 -{3 \over2}\zeta_3\right) 
+ {1 \over 2}C_F T_R N_F + C_A N_F T_R 
\left({2 \over 3} - {\pi^2 \over 9} \right) \; .
\ee

The functions ${\cal C}(Q,b,N,\mu,\mu_F )$ in Eq.~(\ref{crsec})
are given as:
\be
\label{cpdf}
{\mathcal C}_{a/H}(Q,b,N,\mu,\mu_F )
=   \sum_{j,k} C_{a/j}\left(N, \alpha_s(\mu) \right)\,
{\cal E}_{jk} \left(N,Q/\chi,\mu_F\right) \,
               f_{k/H}(N ,\mu_F) \; .
\ee
They are products of parton distribution functions $f_{k/H}$ at scale $\mu_F$,
an evolution matrix ${\cal E}_{jk}$, and coefficients
$C_{a/j}(N,\as)$ which are perturbative series in $\as$.
Explicit expressions for the latter are given in~\cite{KSV,KSVHiggs}.
The matrix ${\cal E} \left(N,Q/\chi,\mu_F\right)$ represents 
the evolution of the parton densities from scale $\mu_F$ to scale 
$Q/\chi$ up to NLL accuracy~\cite{KSV} in $\ln N$.
By incorporating full evolution of parton densities the
cross section~(\ref{crsec}) correctly includes the leading $\as^n
\ln^{2n-1}(\bar N) /N$ collinear non-soft terms to all orders.
Such terms were previously addressed in~\cite{Higgsthreshold}. 
In fact, due to our treatment of evolution, expansion of the resummed cross
section~(\ref{crsec}) in the limit $N \rightarrow \infty,\,b=0$ gives all 
${\cal O}(1/N)$ terms in agreement with the ${\cal O}(\alpha_s)$ result.
Further comparison can be undertaken in the limit $b \rightarrow
\infty,\,N=0$ when our joint resummation turns into standard $Q_T$
resummation. Also, a numerical 
comparison~\cite{KSV,KSVHiggs} between the fixed-order 
and the ${\cal O}(\as)$-expanded jointly resummed expression for 
$d \sigma / d Q_T$ at  shows very good agreement, especially at small
$Q_T$.

\section{INVERSE TRANSFORMS AND MATCHING}

The jointly resummed cross section~(\ref{crsec}) requires defining 
inverse Mellin and Fourier transforms so that 
singularities associated with the Landau pole are avoided. 
A contour for the Mellin integral in~(\ref{crsec}) is chosen in analogy with
the `minimal prescription' contour in threshold resummation~\cite{CMNT}:
\be
\label{cont}
N = C+ z {\rm e}^{\pm i \phi} \; ,
\ee
where the constant $C$ lies to the right of the rightmost singularity 
of the parton distribution functions but left of the Landau pole.

The inverse Fourier integral from $b$ space also suffers from the Landau 
singularity. We define this integral with a similar strategy. We first 
use the identity
\be
\label{bint}
\int d^2 b \;e^{i\vec{q}\cdot {\vec{b}}}\,f(b)\;=\;  2 \pi\,
\int_0^{\infty} \, db\,b\, \,J_0(bq) \,f(b) \;=\; \pi\, 
\int_0^\infty db\, b\, \left[\, h_1(bq,v) + h_2(bq,v)\,\right]\,f(b) \, ,
\ee
and employ Cauchy's theorem to deform the integration over real $b$ 
into a contour in the complex $b$ plane~\cite{LSVprl,KSV}. Here
the auxiliary functions $h_{1,2}$ are related
to Hankel functions. They distinguish
between the positive and negative phases in Eq.~(\ref{bint}). The $b$
integral can thus be written as a sum of two contour integrals, over 
the integrand with $h_1$ ($h_2$) along a contour in the
upper (lower) half of the $b$ plane. The precise form of the contours
becomes unimportant as long as the contours do not run into the Landau pole
or into singularities associated with the particular form~(\ref{chinew}) of 
the function $\chi$. 
Our treatment of contours in complex transform $b$-space is
completely equivalent to the original form, Eq.\ (\ref{bint}), when
the exponent is evaluated to finite order in perturbation theory.
In the presence of the Landau pole arising in the resummed formula, 
it is a natural extension of the $N$-space contour redefinition 
above~\cite{CMNT}, using a generalized ``minimal''
exponent. We emphasize that joint resummation with its contour
integration method  provides an  alternative to
the standard $b$ space resummation.  Joint resummation has built-in
a perturbative treatment of large $b$ values, eliminating the need for a
$b_*$ or other prescription for the exponent, or for a freezing of the
scale of parton distributions at large $b$ or low $Q_T$. In this way,
we can derive entirely perturbative resummed cross sections.

In the joint resummation we adopt the following matching prescription between
the resummed and the fixed-order result:
\begin{equation}
{d \sigma \over d Q^2 d Q_T^2} = {d \sigma^{\rm res} \over d Q^2 d Q_T^2}
-  {d\sigma^{\rm exp(k)} \over d Q^2 d Q_T^2} +
{d \sigma^{\rm fixed(k)} \over d Q^2 d Q_T^2} \,,
\label{joint:match}
\end{equation}
where $d \sigma^{\rm res}/d Q^2 d Q_T^2$ is given in Eq.~(\ref{crsec}) and
$d\sigma^{\rm exp(k)}/d Q^2 d Q_T^2$ denotes the terms resulting from
the expansion of the resummed expression in powers of
$\as(\mu)$ up to the order $k$ at which the fixed-order cross section
$d \sigma^{\rm fixed(k)} /d Q^2 d Q_T^2$  is taken. The above matching prescription
 in $(N,b)$ space guarantees that no double counting of singular contributions
occurs in the matched distribution.

\section{NUMERICAL RESULTS}
\begin{figure}[h]
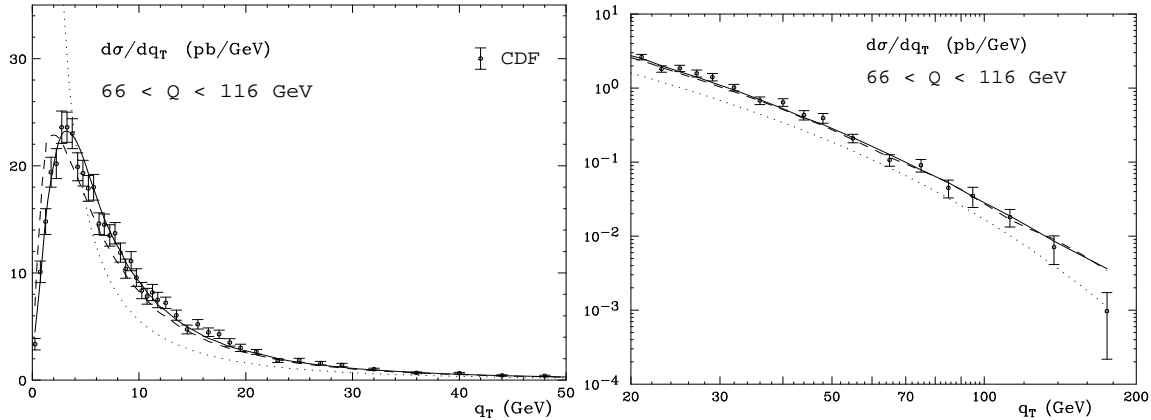

\begin{center}
\includegraphics[width=5.5cm,height=7.5cm,angle=90]{cdf.epsi}
\includegraphics[width=5.5cm,,height=7.5cm,angle=90]{cdf_largeqt.epsi}
\end{center}
\vspace*{-8mm}
\caption{CDF data {\protect \cite{CDF}} on $Z$ production compared to 
joint resummation predictions,
without nonperturbative smearing (dashed) and with Gaussian smearing (solid,
see text). The dotted line shows the fixed-order result. The normalizations
of the curves (factor of 1.035)  have been adjusted in order to give an optimal
description.  We use CTEQ5M~\cite{cteq5m} 
parton distribution functions, $\mu=\mu_F=Q$ and
$\phi=\phi_b=25/32 \pi$, $C=1.3$, $b_c=0.2/Q$.}
\label{fig:cdf}
\end{figure}

\begin{figure}[h]
\vspace*{-4mm}
\begin{center}
\includegraphics[width=9.6cm]{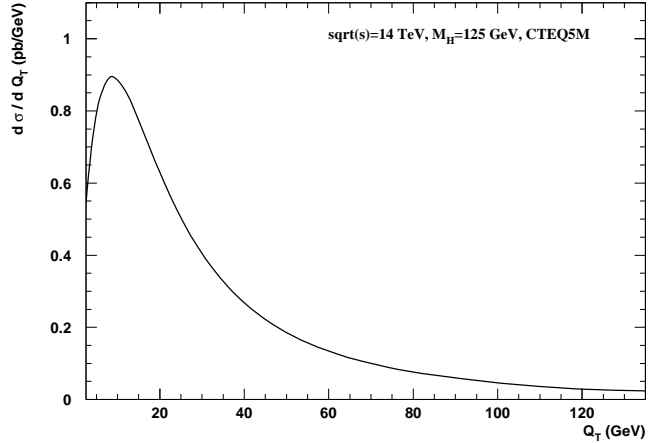}
\vspace*{-8mm}
\caption{Transverse momentum distribution for Higgs production at the LHC in
the framework of joint resummation. We have not implemented any 
nonperturbative smearing. Parton distributions and other parameters
are as in Fig.~1.}
\label{higgs}
\vspace*{-8mm}
\end{center}
\end{figure}

\setcounter{footnote}{0}
Joint resummation predictions for $Z$ boson production compared with the
latest CDF data from the Tevatron collider~\cite{CDF} are shown in
Fig.~\ref{fig:cdf}.  Fig.~\ref{higgs} shows our results for the jointly 
resummed cross section for the production of a 125~GeV Higgs boson at 
the LHC\footnote{Earlier pheomenological studies for the resummed Higgs
production cross section were presented in~\cite{BQ,bcdg}. We note
that the very recent study of~\cite{bcdg} adopts our choice of
contour in complex-$b$ space for the inverse Fourier transform.}.
Due to the contour integral prescription for
performing inverse transforms, in the framework of joint resummation one does
not require any extra nonperturbative information to obtain predictions.
This is not the case in the standard $Q_T$ resummation formalism, where
nonperturbative parameters are introduced to make the theoretical
expression well defined. 

As shown by the dashed line in Fig.~\ref{fig:cdf} the 
joint resummation  without any extra nonperturbative input already provides
a good description of the data for $Z$ production, except for the 
region of very small $Q_T$, 
where the nonperturbative effects are expected to play a significant role.
However, the form of the nonperturbative input can be
predicted within the joint resummation by taking the limit of small
transverse momentum of soft radiation in the exponent, Eq.~(\ref{jointsud}). 
Assuming moderate threshold effects the procedure gives a simple Gaussian
parametrization $F_{NP}(b)=\exp(-g b^2)$. The value of the parameter $g=0.8
\GeV^2$ is determined by
fitting the predicted distribution to the data. It is very similar to the
value obtained in Ref.~\cite{QZ}, where an extrapolation of the exponent to 
large $b$ was carried out for the $Q_T$-resummed cross section. 
The solid line in Fig.~\ref{fig:cdf} represents predictions including the
nonperturbative parametrization. In the large $Q_T$ region, 
see Fig.~\ref{fig:cdf}b, the joint resummation formalism with the matching
prescription~(\ref{joint:match}) also returns a very good description of 
data without requiring an additional switching to a pure fixed-order result, 
unlike in the standard $Q_T$ resummation formalism. 
Nevertheless, at large $Q_T$, no  formalism based on the resummation of 
Sudakov logarithms can be expected to incorporate all relevant
contributions, particularly at small $x$.  The relations between $Q_T$ 
resummation and threshold and joint resummation in the context of Higgs 
production at the LHC should shed light on this issue.


\end{document}